\documentclass{iau}
\usepackage{graphicx}

\title[Photometry Using Kepler ``Superstamps'' of Open Clusters] 
{Photometry Using Kepler ``Superstamps'' of Open Clusters NGC 6791 \& NGC 6819}

\author[Charles A. Kuehn III, Jason Drury, Dennis Stello, \& Timothy R. Bedding]   
{Charles A. Kuehn III$^1$, Jason Drury$^1$, Dennis Stello$^1$, \& Timothy R. Bedding$^1$}

\affiliation{$^1$Sydney Institute for Astronomy, School of Physics, The University of Sydney, NSW 2006, Australia \\ email: {\tt kuehn@physics.usyd.edu.au}}

\pubyear{2013}
\volume{301}  
\pagerange{119--126}
\jname{Precision Asteroseismology: Celebration of the Scientific Opus of Wojtek Dziembowski}
\editors{A.C. Editor, B.D. Editor \& C.E. Editor, eds.}
\begin{document}

\maketitle

\begin{abstract}
The Kepler space telescope has proven to be a gold mine for the study of variable stars.  Unfortunately, Kepler only returns a handful of pixels surrounding each star on the target list, which omits a large number of stars in the Kepler field.  For the open clusters NGC 6791 and NGC 6819, Kepler also reads out larger superstamps which contain complete images of the central region of each cluster.  These cluster images can potentially be used to study additional stars in the open clusters.  We present preliminary results from using traditional photometric techniques to identify and analyze additional variable stars from these images.
\end{abstract}


The high photometric precision and the virtually uninterrupted observing cadence of the Kepler space telescope has made it a revolutionary tool for the study of stellar variability. In order to conserve bandwidth, the spacecraft downlinks a small postage stamp of pixels around each target star, thus photometric data is obtained for only a portion of the stars in the Kepler field.

\begin{figure}
\begin{center}
 \includegraphics[width=0.35\textwidth]{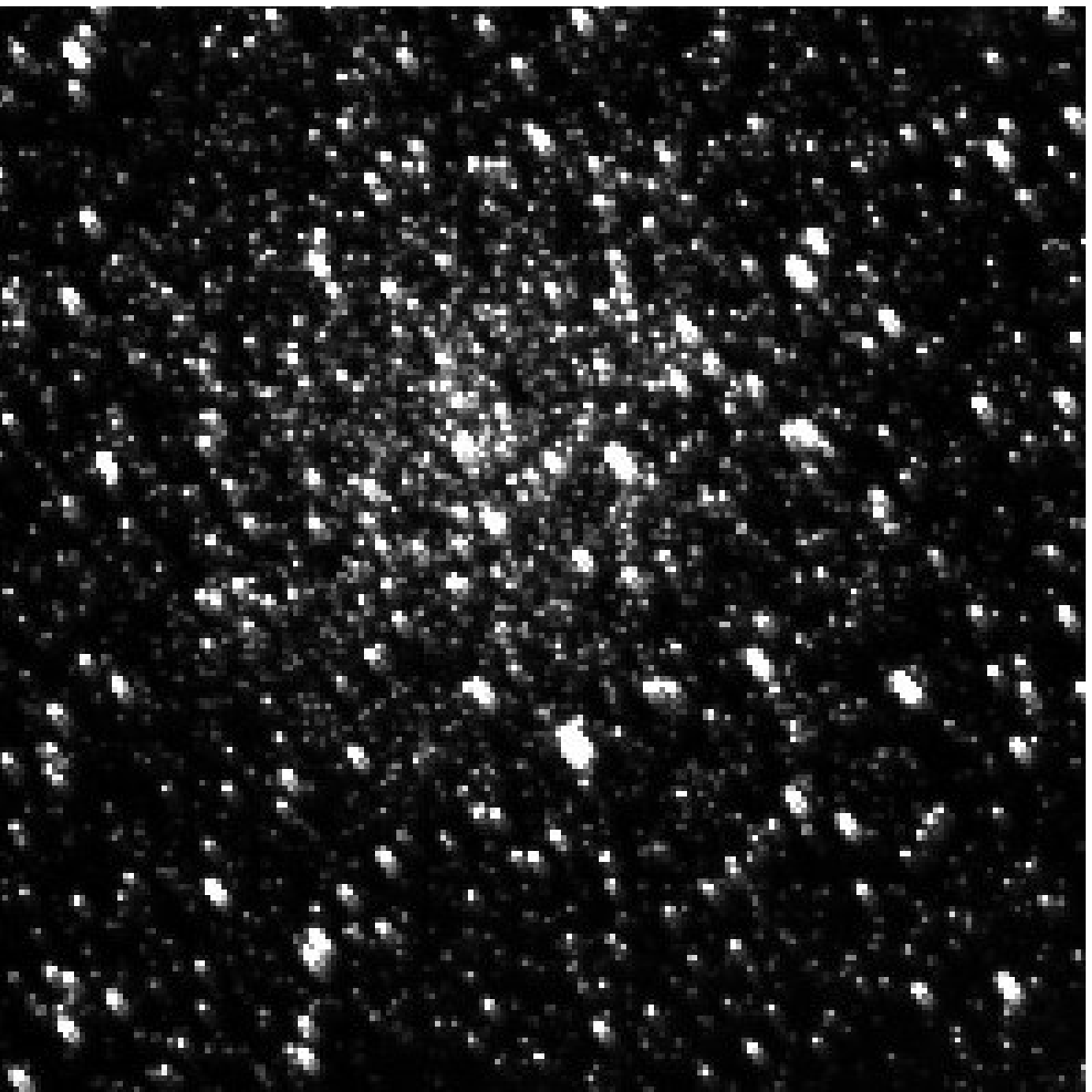}
 \includegraphics[width=0.35\textwidth]{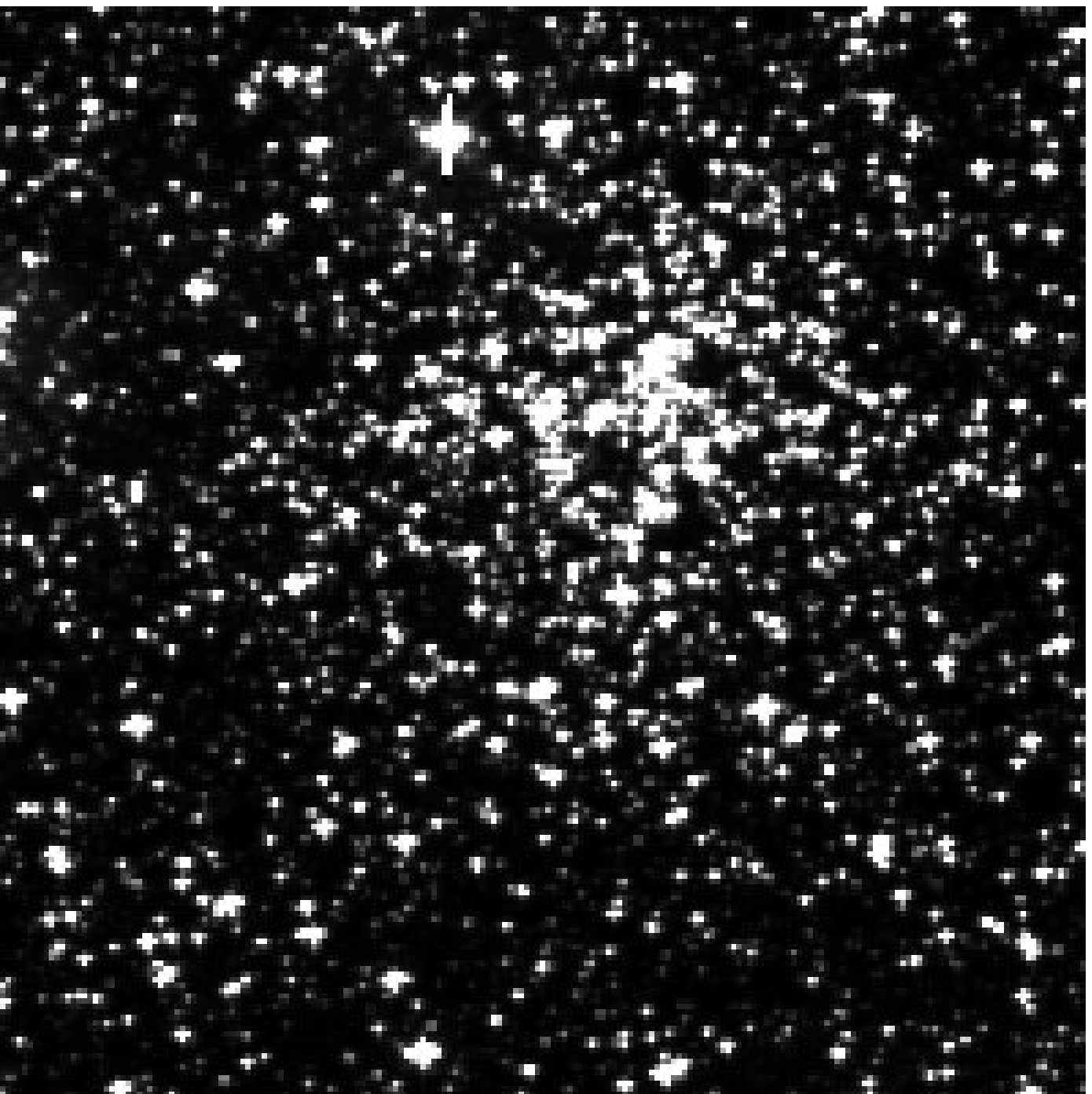} 
 \caption{200x200 pixel superstamp of NGC 6791 (left) and NGC 6819 (right).  Superstamp images created using a routine by Ron Gililand (private communication).}
   \label{fig1}
\end{center}
\end{figure}

Open clusters are ideal targets for asteroseismic studies since all stars in a cluster are thought to have the same age and composition, allowing us to better constrain the stellar models that are compared against the observed oscillation frequencies.  Four open clusters spanning a range of ages and metallicities are located in the Kepler field of view (NGC 6791, NGC 6811, NGC 6819, and NGC 6866).  Target stars were selected and observed in each of the open clusters but the majority of the cluster stars were not targeted.  Results from these observations can be found in \cite[Corsaro et al.(2012)]{corsaro12} and references therein.

In addition to the postage stamps that were used for the work cited above, large $200\times200$ pixel ($13.3$ arcminutes on a side) superstamps of the clusters NGC 6791 and NGC 6819 were also obtained by Kepler in long cadence mode (Figure 1).  These superstamps cover the central regions of the two clusters, providing an opportunity to obtain photometric information on the non-target stars in these regions. The goal of this project is to use traditional photometric techniques to obtain light curves for all of the resolved stars in the superstamps.

The coarse pixel scale of Kepler ($3.98$ arcseconds/pixel) and the crowded nature of the clusters results in many of the stars being blended.  The degree of blending prevents the use of aperture photometry on most of the cluster stars, however profile-fitting photometry is ideally suited for dealing with stars that are partially blended.  Stars were identified in each image and photometry was obtained using Peter Stetson's Daophot/Allstar routines (\cite[Stetson 1994]{stetson94}).  The profiles used for the profile fitting were derived from the actual PSFs of bright, relatively uncrowded stars.  Figure 2 shows the light curves for two stars using the data from quarter 1 of Kepler operation (12 May 2009-14 June 2009).  While the variablity is easily evident in the light curves, there is still a large amount of noise and potential systematics that need to be removed.  Much of the noise is probably from variations in the inter-pixel sensitivies.  Correcting for this requires varying the PSFs based on the location of the star on the CCD; we are currently working on implementing a fitting routine that can do this.

\begin{figure}
\begin{center}
 \includegraphics[angle=270,width=0.35\textwidth]{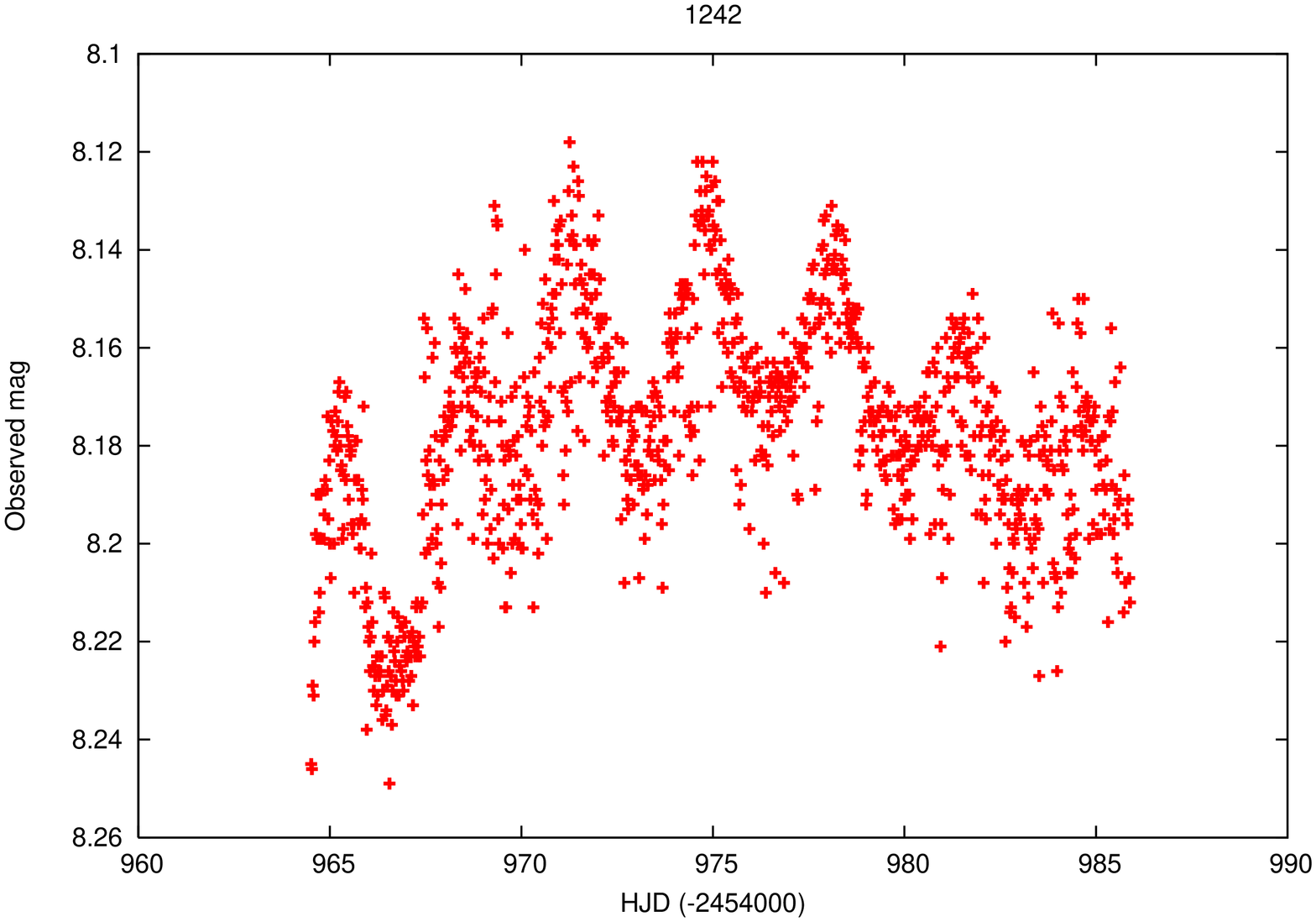} 
 \includegraphics[angle=270,width=0.35\textwidth]{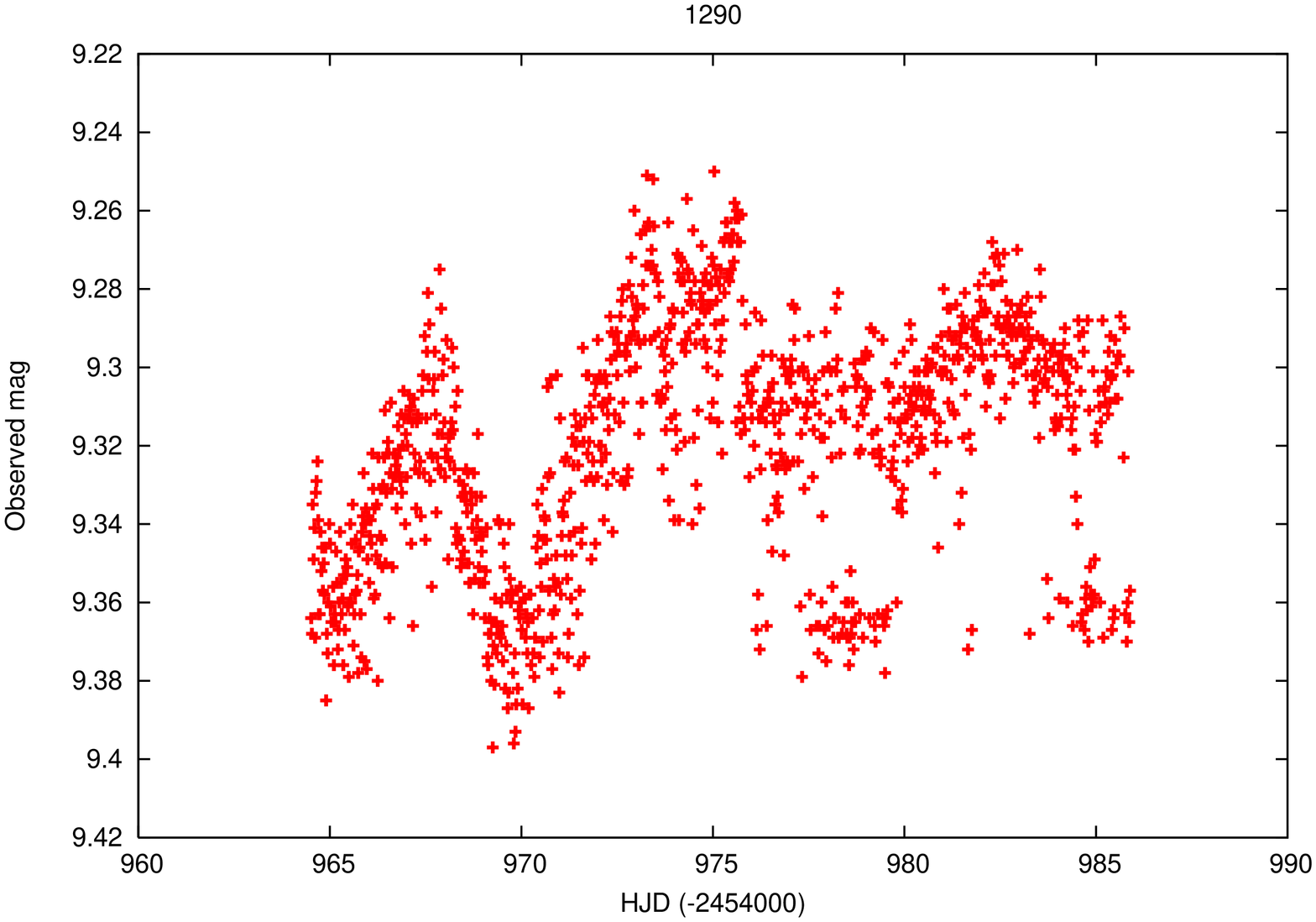} 
  \caption{Sample quarter 1 light curves, obtained with profile-fitting photometry, of stars in NGC 6791.  Magnitudes are given in the instrumental system and have not been converted to the standard system.}
   \label{fig2}
\end{center}
\end{figure}

Image subtraction was also run on the superstamps in order to obtain photometry for stars that were too highly blended for profile fitting.  Wojtek Pych's DIAPL2 package\footnote{http://users.camk.edu.pl/pych/DIAPL/ ; DIAPL2 is an improved version of the DIA package (\cite[Wo$\acute{z}$niak(2000)]{wozniak00})} was used to perform the image subtraction and obtain light curves.  As with the profile-fitting, light curves obtained from the image subtraction method show a good deal of noise as well as systematic effects.  We are currently in the process of refining the parameters of the image subtraction process to decrease the noise and systematic effects.

The preliminary results show that it is possible to obtain light curves from the Kepler superstamps of the open clusters NGC 6791 and NGC 6819, allowing the study of the variability of stars not included in the Kepler target list.

\end{document}